\documentstyle[prl,aps,epsfig,floats,amsfonts]{revtex}   
\def \cl {\centerline}
\def \be {\begin{equation}}
\def \ee {\end{equation}}
\def \d  {\delta}
\def \jpa {J. Phys. A}
\def \la {\langle}
\def \ra {\rangle}

\begin{document}
\twocolumn \psfull \draft 
\wideabs{   
\title{Generic Sandpile Models Have Directed Percolation Exponents}
\author{ P. K. Mohanty  and Deepak Dhar} 
\address{Tata Institute of Fundamental Research, Homi Bhabha Road, 
Mumbai-400 005, INDIA}

\maketitle 
\begin{abstract} 

We study sandpile models with stochastic toppling rules and having sticky
grains so that with a non-zero probability no toppling occurs, even if the
local height of pile exceeds the threshold value. Dissipation is
introduced by adding a small probability of particle loss at each
toppling.  Generically, for models with a preferred direction, the avalanche
exponents are those of critical directed percolation clusters.  For undirected 
models, avalanche exponents are those of directed percolation clusters in
one higher dimension.

\end{abstract}
} 

  In recent years, there has been a lot of interest in the study sandpile
models as models of real granular matter \cite{granular}, and also as
paradigms of self-organized critical systems in general\cite{soc}.
Following the well-known work of Bak, Tang and Wiesenfeld\cite{btw}, many
different types of sandpile models with different toppling rules have been
studied \cite{kadanoff} : deterministic and stochastic, with or without
preferred direction, different instability criteria \cite{manna1}, or
particle distribution rules \cite{maslov}, with fixed energy
\cite{vespignani} etc.. However, a clear picture of the factors that
determine different universality classes of critical behavior is yet to
emerge \cite{biham}.

  A different paradigm for non-equilibrium critical phenomena has been
directed percolation (DP) which is believed to describe active to
absorbing state transition in a wide class of reaction-diffusion systems
\cite{hinrichsen}. The activity of avalanches in sandpile models can grow,
diffuse or die, and any stable configuration is an absorbing state. Thus,
this should be in the universality class of DP with many absorbing states
\cite{munoz}.  Several models of self-organized criticality which show
critical exponents related to DP have been studied earlier\cite{sneppen}.
However, these models do not involve any conserved field.  The critical
exponents of known models with conservation of sand are very different
from those of DP, and this is presumably due to role of the local
conservation of sand in the model.  In \cite{maslov}, a model with
conservation of particles showing DP exponents was studied, but this study
was mainly numerical.

In this Letter we study several sandpile models with stochastic toppling
rules, both directed and undirected. The grains are `sticky' in our models
in the sense that there is a nonzero probability that any grains arriving
at a site during the avalanche process just get stuck there. We find that
the distribution function of avalanches in these models has the same power
law distribution as that of the critical DP clusters. Our theoretical
arguments, supported by numerical simulations, show that generically 
these models  belong to the DP universality class. The previously 
studied deterministic models \cite{btw,dr}, and the stochastic toppling 
model introduced by Manna \cite{manna} are unstable to perturbations and
flow to the DP fixed point.

  The relation of sandpile models to DP was attempted earlier in
\cite{tadic} using sticky grains. However, in that paper, no bulk
dissipation was present, and sand was dissipated only at the boundaries.
The boundaries break the translational invariance of the steady state. The
density becomes space dependent, and the density profile affects the
statistics of avalanches. The critical exponents of avalanches are not
those of critical DP clusters, but expressible in terms of DP exponents
\cite{note1}. Introducing bulk dissipation in this paper, we are able to get
rid of these problems, and the relationship between the sandpile and DP
problems becomes more transparent.

For simplicity, we start by defining the directed model on a
$(1+1)$-dimensional square lattice. Various generalizations to higher
dimensions, undirected case, and other toppling rules are straight
forward, and briefly discussed at the end.  The sites on an $L \times M$
torus are labelled by euclidean coordinates $(i,j)$ with $(i+j)$ even and
$j$ increasing downwards. At
each site $(i,j)$, there is a non-negative integer $h_{i,j}$ to be called
the height of the pile at that site. Initially all $h_{i,j}$ are zero.  
The system is driven by choosing a site at random and increasing the
height at that site by one. If one or more particles are added to a site
at time $t$ (from outside or from other sites), and its height becomes
greater than $1$, then it is said to become unstable at time $t$.

Any site $(i,j)$ made unstable at time $t$ relaxes at the time $(t+1)$
stochastically: With probability $( 1-p)$, it becomes stable without
losing any grains. We say that the added particle(s) sticks to the
existing grains.  Otherwise (with probability $p$), the relaxation occurs
by toppling in which the height at the site decreases by two, and the site
{\it becomes stable}. We introduce bulk dissipation, by assuming that in
each toppling, there is a small probability $\delta$ that both grains from
the toppling are lost, other wise (with probability $1-\delta$), the two
grains are transferred to the two downwards neighbors $(i \pm 1,j+1)$.

Note that there is a nonzero probability that a stable site can have
arbitrarily large heights. We relax all the unstable sites by parallel
dynamics.  A site made unstable at time $t$ is relaxed in one step at
time $t+1$, independent of whether it received one or more grains at the
previous time step. Once a site has relaxed, it remains stable until
perturbed again by new grains coming to the site. This relaxation process
is repeated until all sites become stable, and then a new grain is added.

The model is specified by two parameters $p$ and $\delta$. The case $p=1,
\d=0$ is exactly soluble, and its critical exponents are known in all
dimensions \cite{dr}. In this case, one has to introduce open boundary
conditions to ensure the existence of a steady state. In two dimensions,
the probability that adding a particle will cause an avalanche of s
topplings varies as $s^{-\tau_s}$ for large $s$ with $\tau_s = 4/3$. The
probability that the duration of avalanche equals $T$ varies as
$T^{-\tau_t}$ with $\tau_t =3/2$.  The case $\d=0, p$ arbitrary was
studied earlier in \cite{tadic} discussed in the introduction.
\epsfxsize=3.3 in
\epsfysize=2.2 in
\cl{\epsffile{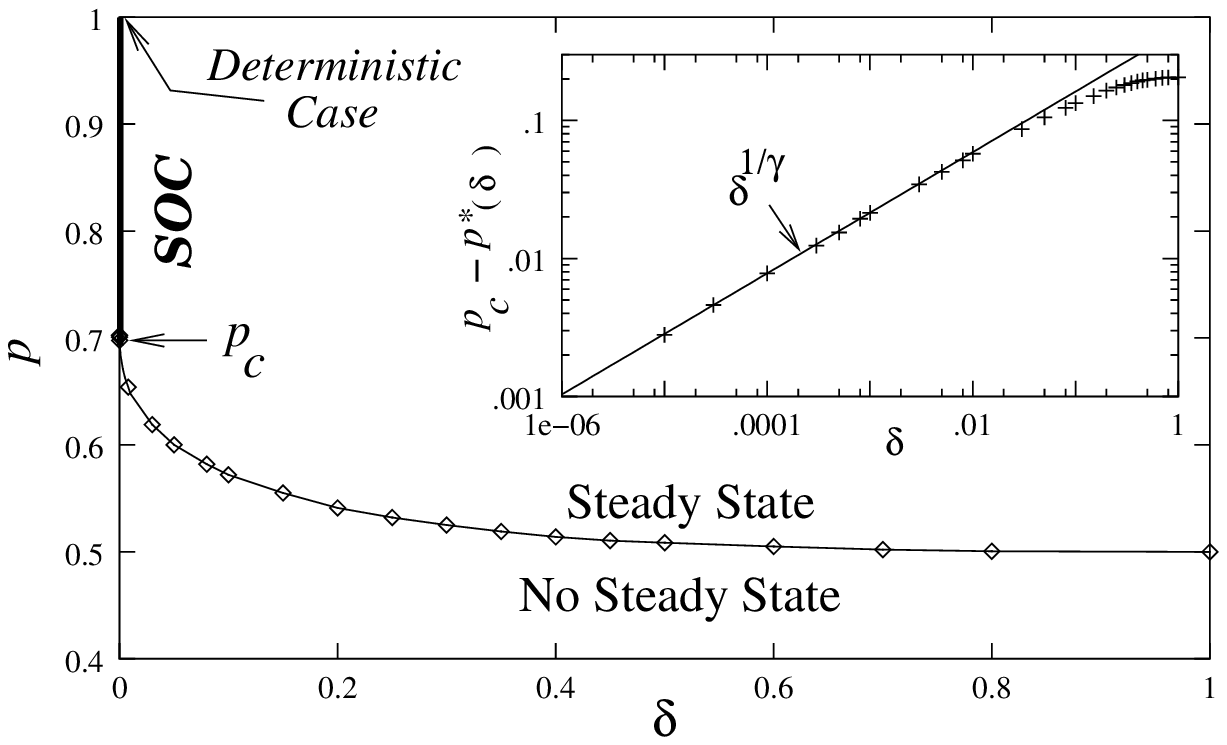}}
{\bf Fig. 1} The line $ p = p^*(\d)$ separating the steady state and
no-steady-state regions in the $p$-$\d$ plane. The inset shows a plot of
$[p_c-p^*(\d)]$ versus $\d$. The straight line shows the theoretical slope
$1/\gamma$. 

The probability distribution of different configurations in the steady
state for general values of $p$ and $\d$ have an interesting structure.
Define a variable $g_{i,j} = h_{i,j} ( mod~ 2)$.  We group together
different stable configurations $\{h_{i,j}\}$ corresponding to the same
values of $\{g_{i,j}\}$. There are $2^{LM}$ such equivalence classes. On
addition of a particle to a site, the $g$ variable at that site flips. A
toppling leads to flipping of $g$'s at the two downward sites. From
detailed balance, it follows that in the steady state, each of the
$2^{LM}$ equivalent classes occurs with equal probability \cite{chhabra}.

If $p=1$, with $\d$ arbitrary, the only allowed height values are $0$ and
$1$. In this case, we get a full characterization of the steady state.  
The $n$-point correlation functions satisfy linear equations, and exact
solution of \cite{dr} can be generalized to arbitrary $\d$.  We omit the
details here. The distribution of avalanche-sizes has an exponential decay
for non-zero $\d$.

As $p$ is decreased below $1$, heights greater than $1$ appear with
non-zero probability in the steady state. The mean height $\la h_{i,j}\ra$
increases as $p$ is decreased, and there exists a critical value $p^*(\d)$
such that for $p \leq p^*(\d)$, the height of the pile increases without
bound, and there is no steady state (Fig. 1).

The absence of a steady state is obvious along the the line $p=0$, with
$\d$ arbitrary, as in this case, different sites cannot influence each
other, and each added particle just sticks to the existing pile. A similar
decoupling occurs along the line $\d =1$, with $p$ arbitrary. In this
case, the average particle loss per added grain is $ 2( 1 - c_0) p$, where
$c_0$ is the density of sites with height $0$. In the steady state this
must equal $1$. As $c_0 \geq 0$, a steady state can exist only for $ p >
1/2$. Thus $p^*(\d=1) = 1/2$.

We now derive the equation for the boundary line $p = p^*(\d)$. At the
phase boundary, clearly $c_0 = 0$. In the growth of an avalanche in the
steady state of the system, any site which receives at least one grain
from its upward neighbors sends grains to downward neighbors with a
probability $p(1-\d)$.  Thus the probability $Prob(s)$ that an added grain
will cause an avalanche in which at least $s$ sites transfer particles to
downward neighbors is {\it same as} the probability
$Prob_{DP}(s|\tilde{p})$ of a cluster of at least $s$ sites in a {\em
directed site percolation process} with concentration of active sites$=
\tilde{p}$. A \begin{equation} Prob(s) = Prob_{DP}(s ~|~ \tilde{p}=
p(1-\d)). \end{equation} Note that Eq.(1) also holds in the entire `no
steady state' phase, where mean height continues to increase, but
$Prob(s)$ tends to a limiting distribution for large times.

By particle conservation, in the steady state, the average number of
topplings in an avalanche must be equal to $1/(2 \d)$. 
Let $n_{_{DP}}(\tilde{p})$ is the sum of the average number of occupied
and perimeter sites in the cluster corresponding to a randomly picked site
in the DP problem. Now, the equation for the phase boundary  
can be expressed in terms of the function
$n_{_{DP}}(\tilde{p})$ as
\begin{equation}
 n_{_{DP}}( (1-\d)p^*(\d)) =1/(2 \d).
\end{equation}

For small $\d$, the average size of clusters is large, and $\tilde{p}$ is 
near the critical probability $p_c$ for the directed site percolation on
this lattice, the function $n_{DP}$
is known to vary as $(p_c - \tilde{p})^{-\gamma}$. Substituting this in
Eq.(2) we get

\begin{equation}
p^*(\d) =  p_c - A \d^{1/\gamma} +{\rm terms ~of ~higher ~order ~in~} \d,
\end{equation}
where $A$ is some constant, and $\gamma$ is the susceptibility exponent of
the DP problem. In particular, we have $p^*(\d=0) = p_c$. The inset in
Fig. 1 shows a log-log plot of the numerically determined values of $p_c -
p^*(\d)$ plotted versus $\d$.  We get good agreement with Eq.(3) using the
existing estimates $p_c = 0.70548515$ and $\gamma = 2.277730$
\cite{jensen}.

Since the average number of
topplings in an avalanche in the steady state is $1/(2 \d)$,
we will see self- organized criticality with long-ranged correlations
only in the limit $\d \rightarrow 0^{+}$, and $ p_c
\leq p \leq 1$, (marked with a heavy line in Fig. 1).
	For $\d =0^+ $ and $1 \geq p > p_c$, in the steady state $c_0 >
0$.  Each site is characterized by two probabilities, $p_1$ and $p_2$,
which correspond to the probability that a site topples when one or two
particles are added to it respectively. Clearly, $p_1=(1- c_0)p$ and $
p_2=p$.  The correlations between heights at different sites are small,
and if we ignore them \cite{note}, the evolution of avalanches (Fig. 2) 
in the steady state is as in the Domany-Kinzel model of DP with 
parameters $(p_1, p_2)$ \cite{dk}. 

\epsfxsize=3.0 in
\epsfysize=1.8 in
\cl{\epsffile{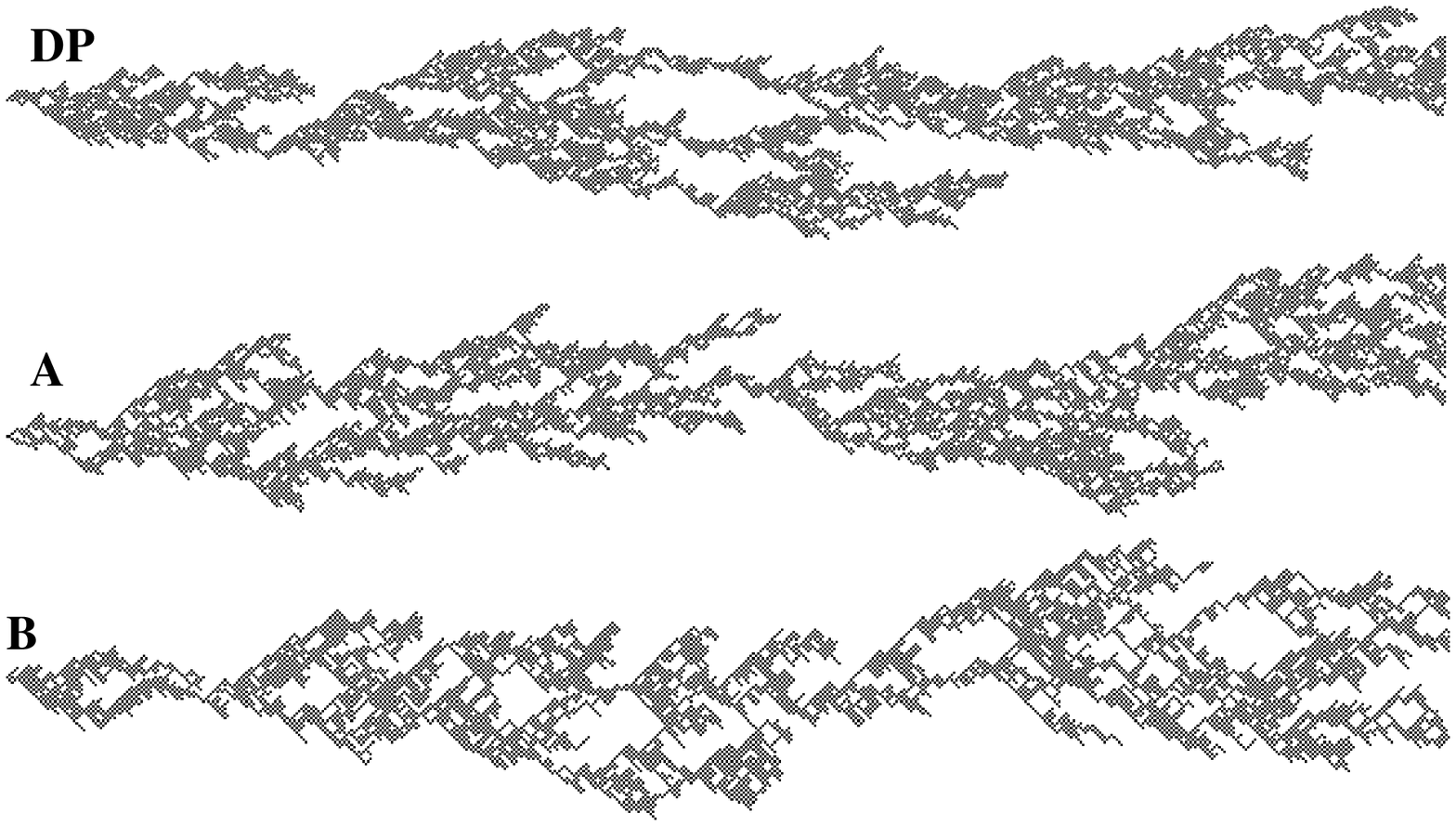}}

{\bf Fig. 2} Picture of a typical avalanche for the 2-d directed model (A)  
and the time-evolution of the 1-d undirected model (B) for $ p =.873 > p_c$
and $\d=.0001$ are compared with the clusters at the critical line of the
Domany-Kinzel model of DP with $p_2 =p$.

Even if some short-range correlations are present, they should not modify
critical behavior, which is expected to be same as in DP. In Fig. $3$, we
have compared the the probability distribution of avalanches with that of
DP clusters. For $s \gg 1$, the latter is expected to satisfy the scaling
form

\be 
Prob(s) \approx \frac{A}{s^{\tau-1}}f[ s/s^{*}], 
\ee 
where $A$ is some amplitudes, $\tau$ is a critical exponent for cluster
size distribution in DP and the function $f(x)$ tends to a finite constant
as $x$ tends to zero, and decreases exponentially with $x$ for large $x$.
We assume that $s^{*}$ varies as $\d^{-\phi}$. Using the constraint $\la
s\ra \sim \d^{-1}$, we get $\phi=1/(2-\tau)$. Using the known numerical
estimate $\tau = 1.108$ in $d=1+1$ \cite{jensen}, we find a very good
collapse when $ s^{\tau -1} Prob(s)$ is plotted versus $s \d^{\phi}$ for
two different values $p = p_c$ and $p=0.873$, and two values of $\d =
10^{-3}$ and $\d = 10^{-4}$. In the inset, the scaling function $f$ is
compared with that for DP clusters. We get an excellent collapse,
a strong evidence that the two functions are {\it the same}, and the
correlations in heights in the steady state  are irrelevant.

It is straightforward to extend the previous discussion to higher
dimensions.  Thus the avalanche exponents in the $(d+1)$-dimensional model
are the same as the exponents of cluster size distribution in the
$(d+1)$-dimensional DP at critical point. The upper critical dimension is
$d=4$.

Consider now the undirected version of the problem on a $d$-dimensional
hypercubical lattice. The rules are the same as before, except that on
toppling, a particle is transferred to each of the $2d$ neighbors of the
toppling site. Clearly, in this case also, there is no steady state for
small $p$. For $p= p^*(\d)$, the mean height per site diverges. Then, any
site which has one or more particles added to it, sends particles to its
neighbors with probability $p(1-\d)$. 
If we look at the space-time history
of the evolution of the avalanche, we get a directed site-percolation
cluster on a $(d+1)$-dimensional body centered hypercubical (bch) lattice.
The phase boundary in this undirected model is also the same as that of the
directed model on the $(d+1)$-dimensional bch lattice.

 \epsfxsize=3.3 in
 \epsfysize=2.2 in
 \cl{\epsffile{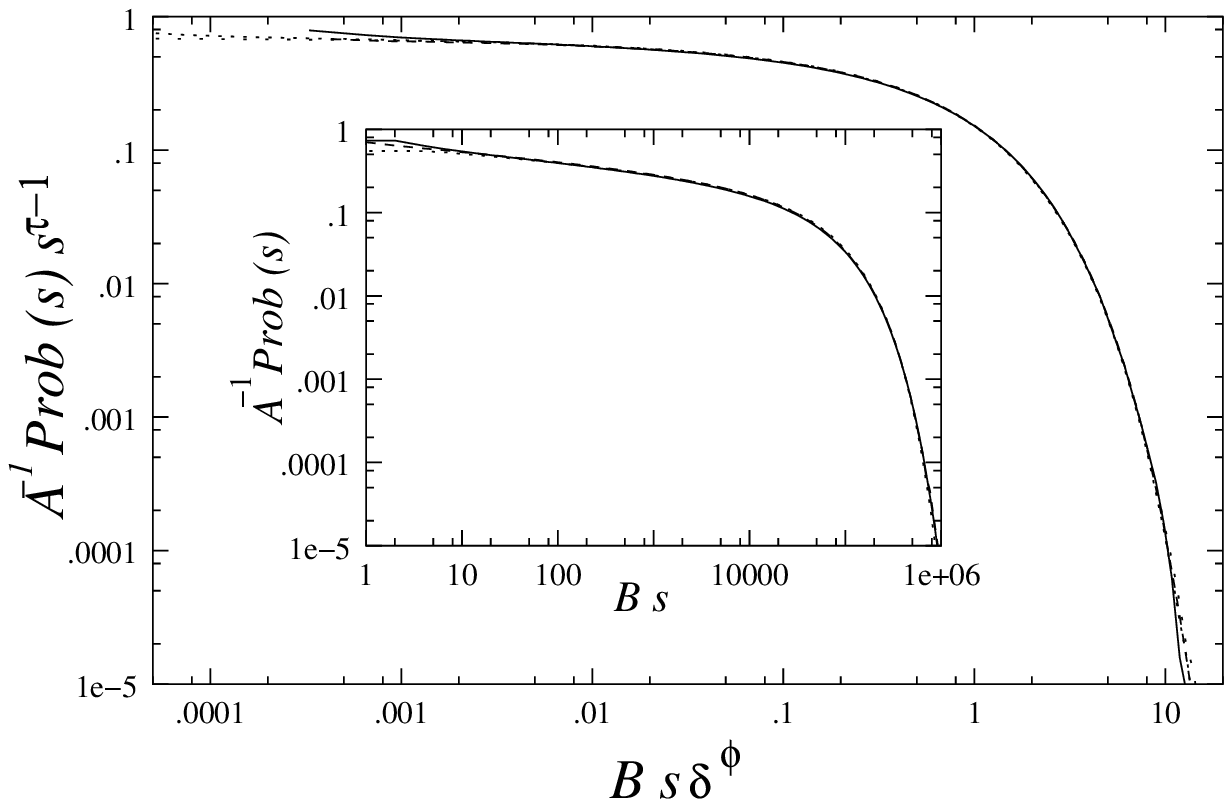}}

{\bf Fig. 3} Scaling collapse of $ s^{\tau -1}
{\rm Prob}(s) $ versus $s\d^\phi$ for four different combinations of
parameters $p,\d$. The inset compares the scaling function for the 
directed and undirected models with $(p,\d)=(0.873,10^{-4})$ with that for
DP at $p=0.7$.

For $p > p^*(\d)$, the undirected model differs from the directed model in
that the height of the pile at a site does not change between two
topplings. This may give rise to possible long term memory effects. However, 
in our model this effect is rather small, as the probability of toppling
depends on height only if the height is zero. Along the line $p=p^*(\d)$,
the memory affect is strictly absent, as the density of sites with zero
height goes to zero. We find that even for $p$ as large as $.873$, the
avalanches are qualitatively similar to DP (Fig. 2),  and the distribution  
function is also indistinguishable from that of near-critical DP clusters 
(see inset of Fig. 3). There is a crossover from deterministic limit ($p=1,
\d=0$) \cite{btw} to DP for $p\ne 1$.

Why does the conservation of particles not change the critical behavior
away from DP in our problem? Consider a simple DP process of particles of
type X, on a lattice with $n_X(\vec{r})$ particles of X at site $\vec{r}$.
We now attach a register $n_Y(\vec{r})$ at each site $\vec{r}$, which
decreases (increases) by 1 each time a particle X is created ( destroyed)
at $\vec{r}$. Then clearly, we have a local conservation of $n_X + n_Y$.
Clearly, if the dynamics of the X particle is not affected by the book
keeping, the process still belongs to the DP universality class. In our
model, $n_Y$ is the height of the pile. It fluctuates about its mean
value, but there is an influence of $n_Y$ on the dynamics of X particles
as birth of X particles is not allowed if $n_Y$ is zero.

The phenomenological
evolutions for the coarse-grained density fields  $n_X(\vec{r})$ and
$n_Y(\vec{r})$ in the conservative limit $\d = 0^+$ may be written as 
\cite{vespignani2}
\begin{eqnarray}
\partial_t  n_X =  \bigtriangledown^2 n_X+ a n_X&&-b
n_X^2  + c n_X g(n_Y) + \eta( \vec{r},t) \\
\partial_t(n_X+n_Y) =  \bigtriangledown^2 n_X&&
 \end{eqnarray}
where $\eta(\vec{r},t)$ is a noise term, and $a, b$ 6and $c$ are
phenomenological constants. The crucial term here is the coupling term 
$c n_X g( n_Y)$. Vespignani et al \cite{vespignani2} chose $g(n_Y)$
proportional to
$n_Y$. In our case, $n_Y$ has a threshold and the effect on $n_X$
saturates to a finite value even as $n_Y$ increases to infinity.  The
simplest choice of $g(n_Y)$ to model this is to choose 
$g(n_Y)=\theta(n_Y- h_c)$, where $\theta$ is the step function and $h_c$
is 
the threshold value.   
In naive power counting, this term has the same scaling as the linear term
in $n_X$. We are not able to treat this analytically. But the results of
our simulations strongly indicate that this perturbation does not change
the critical exponents \cite{notepaczusky}.

\epsfxsize=2.2 in
\epsfysize=1.1 in
\cl{\epsffile{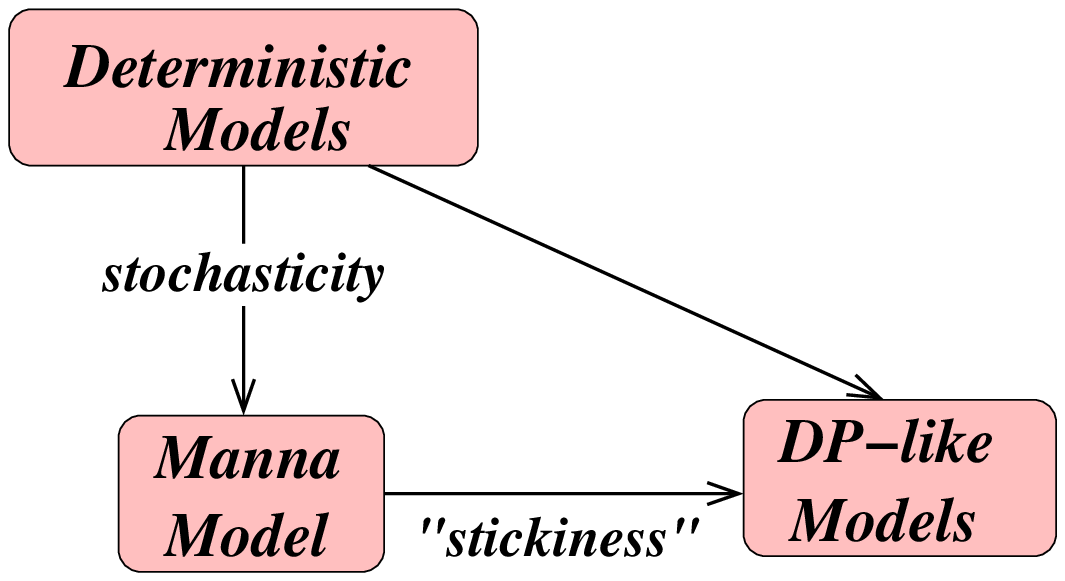}}
{\bf Fig. 4} A schematic flow diagram of renormalization group flows
between different fixed points of sandpile models.

The DP fixed point is expected to be rather robust against perturbations.
We have tested several variations of toppling rules in simulations. One
can make the particle transfer process stochastic with each transferred
particle going to a randomly chosen downward neighbor or one can allow
multiple topplings at a site with each site toppling twice, thrice, etc.,
with decreasing probability, so long as the height is $> 1$. With both
multiple topplings and stochasticity in particle transfer, in the limit of
no stickiness this becomes Manna model \cite{manna}. For sticky grains, in
all these models, we get the DP behavior. The schematic renormalization
group flows are shown in Fig. 4.

 To summarize,  we have studied several sandpile models which show 
DP-like behaviour. A feature, which is common in all these models is 
{\em `stickiness'}, $i.e.$ with a nonzero probability a site can remain 
stable even with a height greater than the threshold. This behavior seems to
be robust against perturbations, and is the generic behavior of sandpile
models, both directed, and undirected. 

 We thank M. Barma, R. Dickman and A. Vespignani for their comments on the
manuscript.


\begin{references}
\bibitem{note} Note that in the steady state, the heights are
independently distributed along the three boundaries $p=0$, $p=1$ and
$\d=1$.
\end{references}
\end{document}